\documentclass{iopart}
\newcommand{\rb}{{\bi{r}}}             
\newcommand{\ab}{{\bi{a}}}
\newcommand{\bb}{{\bi{b}}}
\newcommand{\hb}{{\bi{h}}}

\newcommand{\Rb}{{\bi{R}}}
\newcommand{\Ri}{{{\bi{R}}_i}}

\newcommand{\tW}{\tilde{W}}
\newcommand{\tRi}{{\tilde{\bi{R}_i}}}

\newcommand{\Ec}{{\mathcal{E}}}

\newcommand{\mOmega}{{\mathit{\Omega}}}
\newcommand{\erfc}{\mathop{\rm{erfc}}\nolimits}
\newcommand{\sump}{\mathop{{\sum}'}}

\begin{document}
\title[Convergence peculiarities of lattice summation upon multiple charge 
spreading]
{Convergence peculiarities of lattice summation upon multiple charge 
spreading generalizing the Bertaut approach}
\author{Eugene V Kholopov}
\address{A V Nikolaev Institute of Inorganic Chemistry, Siberian Branch,
Russian Academy of Sciences, 630090 Novosibirsk, Russia}
\address{Novosibirsk State University, 630090 Novosibirsk, Russia}
\ead{kholopov@che.nsk.su}
\begin{abstract}
Within investigating the multiple charge spreading generalizing the Bertaut 
approach, a set of confined spreading functions with a polynomial behaviour,
but defined so as to enhance the rate of convergence of Coulomb series even
upon a single spreading, is proposed. It is shown that multiple spreading
is ultimately effective especially in the case when the spreading functions of
neighbouring point charges overlap. In the cases of a simple exponential and 
a Gaussian spreading functions the effect of multiplicity of spreading on
the rate of convergence is discussed along with an additional optimization of
the spreading parameter in dependence on the cut-off parameters of lattice
summation. All the effects are demonstrated on a simple model NaCl structure. 
\end{abstract}

\section{Introduction}
It is known that the Ewald approach \cite{Ewal21} is the most
widespread implementation of the Poisson summation formula in
crystals \cite{Kawa99,Kawa00,Khol04,Khol07}. Bearing in mind that 
such a treatment proposes that the overall summation is divided 
into sums over direct and reciprocal space, the enhancement of 
the rate of convergence is traditional in this problem and assumes 
the investigation of ranges of summation in either of those sums. 
Indeed, Epstein \cite{Epst03} discussing this problem for the first 
time has proposed that both the sums must be separated by a dimensionless
parameter unity, in agreement with conventional mathematical
approaches. According to Ewald \cite{Ewal21}, the corresponding 
parameter of splitting can still be chosen as a unique, but variable, 
so as to provide the most rapid rate of convergence of both the sums. 
This proposal remained appreciable for a long time \cite{Dien48}. 
However, in the last years one more standpoint arises, keeping in 
mind that the splitting parameter can also depend on the cut-off 
parameters of both the summations in question \cite{Raja94,Humm95}.
In particular, such a treatment can reduce the computational efforts 
associated with the dimension of a supercell employed in molecular 
dynamics \cite{Perr88,Whee02}.

It is important that the Ewald approach can be regarded as an
effect of charge spreading with a Gaussian spreading function
\cite{Ewal21,Khol04,Whee02}. In this connection, the Bertaut treatment
\cite{Bert52} is obviously the extension of the Ewald one to an 
arbitrary spreading function. It implies that the problems of 
adjustable parameters of splitting still remain actual in this 
generalized treatment as well \cite{Raja94,Heye81}. Of course, 
these problems become immaterial if the spreading functions applied 
to different point charges in a point-charge lattice do not overlap 
\cite{Bert52,Bert78}. In this case the sum over direct space is absent 
and so the error arising upon truncating the summation over reciprocal 
space is the only subject of interest \cite{Temp55,Jone56,Jenk71,%
Herz79,Jenk79}. 

It is worth noting that the charge spreading with a certain spreading 
function applied to all couples of interacting charges in the lattice 
was originally discussed \cite{Bert52,Bert78}. On the other hand, it 
turns out that spreading the charges generating the potential field is 
sufficient for enhancing the rate of convergence of the lattice series 
\cite{Heye81,Ween75,Argy92}. Nevertheless, the application of spreading
to all the charges in the expression for the Coulomb energy appears
to be somewhat more efficient \cite{Bert78}. The reason of this
efficiency can be understood from the fact that, by symmetry, the latter 
effect may be treated as a double spreading of charges generating the
potential field. As a result, the idea of a multiple charge spreading
arises as a next step towards achieving the faster convergence
\cite{Kho108}.  

In the present paper the problem of enhancing the rate of convergence 
is discussed in detail. We consider different principal classes of 
spreading functions, with concentrating our attention on the effects of 
multiple spreading. As particular cases of spreading functions extended 
to infinity, here we discuss a Gaussian spreading function and a simple 
exponential spreading. Furthermore, we examine different types of confined 
spreading functions in a sequence of enhancing their convergence efficiency, 
providing that the possibility of their overlap is accessible. To our mind, 
the latter effect extends the original ideas of Bertaut \cite{Bert52,Bert78}.

\section{Basic relations describing the multiple charge spreading}
For convenience, here we compile some results describing the multiple charge 
spreading and obtained earlier \cite{Kho108}. Every perfect crystal can be 
specified by a unit-cell charge distribution $\rho(\rb)$ subject to the 
condition of neutrality of a unit cell :
\begin{equation}\label{Kq1}
\int_V\rho(\rb)\,d\rb=0 ,
\end{equation}
where the integral is over the volume occupied by $\rho(\rb)$. The corresponding
structure factor as a function of a reciprocal lattice vector $\hb$ is of the 
form
\begin{equation}\label{Kq2}
F(\hb)=\frac{1}{v}\int_V\rho(\rb)\exp(-2\pi i \hb\rb)\,d\rb ,
\end{equation}
where $v$ is the unit cell volume. Then the relation $F(\hb=0)=0$ readily follows 
from (\ref{Kq1}). In the particular case of a point-charge lattice the charge
distribution is converted into
\begin{equation}\label{Kq3}
\rho(\rb)=\sum_j q_j\delta(\rb-\bb_j) ,
\end{equation}
where $j$ runs over point charges $q_j$ belonging to a unit cell and located 
at positions $\bb_j$, $\delta(\rb)$ is the Dirac delta function.

The effect of charge spreading will be described by a spherically symmetric
spreading function $\sigma(|\rb|)$ normalized by the condition
\begin{equation}\label{Kq4}
\int\sigma(|\rb|)\,d\rb=4\pi\int_0^\infty\sigma(r)r^2\,dr=1 .
\end{equation}
Here $r=|\rb|$. The application of $\sigma(|\rb|)$ to the charge distribution 
in a unit cell results in an additional multiplier
\begin{equation}\label{Kq5}
S(\hb)=\int\sigma(|\rb|)\exp(-2\pi i \hb\rb)\,d\rb 
\end{equation}
modifying structure factor (\ref{Kq2}) as follows:
\begin{equation}\label{Kq6}
F(\hb)\to F(\hb)S^n(\hb) 
\end{equation}
if the spreading of interest is performed $n$ times in a consecutive manner.

The corresponding potential effect of a multiple spreading in direct space
is described by the function
\begin{equation}\label{Kq7}
\mOmega^{(n)}(|\Rb|)=\int\frac{\sigma(|\rb_1|)\dots\sigma(|\rb_n|)\,
d\rb_1\dots d\rb_n}{|\Rb+\rb_1+\dots+\rb_n|} .
\end{equation}
As a result, it is found that the electrostatic potential field arising 
within such a procedure of spreading can be written as
\begin{eqnarray}\label{Kq8}
U_{(n)}(\rb)&=&\frac{1}{\pi}\sump_\hb\frac{F(\hb)S^n(\hb)}{|\hb|^2}
\exp(2\pi i\hb\rb)\nonumber\\
&&{}+\sump_i\int_Vd\rb_1\rho(\rb_1)\frac{W^{(n)}(|\tRi|)}{|\tRi|}
-\Bigl\{q_j\mOmega^{(n)}(0)\Bigr\}_{\rb=\bb_j} ,
\end{eqnarray}
where the prime on the summation sigh over reciprocal lattice vectors $\hb$ 
implies missing the term at $\hb=0$, the parameter $i$ runs over vectors
$\Ri$ determining the Bravais lattice and specifying different unit cells, 
we introduce the notation $\tRi=\Ri+\rb_1-\rb$, the prime on the 
summation sigh over $i$ means that a possible singularity associated with 
the denominator in the summand must be omitted as well,
\begin{equation}\label{Kq9}
\frac{W^{(n)}(R)}{R}=\frac{1}{R}-\mOmega^{(n)}(R) ,
\end{equation}
the last term on the right-hand side of (\ref{Kq8}) is the correction 
associated with a point charge $q_j$ if it happens at $\rb$. Taking 
potential (\ref{Kq8}) into account, we write down the Coulomb energy 
per unit cell in the form 
\begin{eqnarray}\label{Kq10}
\Ec_{(n)}&=&\frac{v}{2\pi}\sump_\hb\frac{|F(\hb)|^2S^n(\hb)}{|\hb|^2}
+\frac{1}{2}\sump_i\int_Vd\rb_1\,d\rb_2\rho(\rb_1)\rho(\rb_2)\nonumber\\
&&{}\times\frac{W^{(n)}(|\tRi_{12}|)}{|\tRi_{12}|}-\frac{\mOmega^{(n)}(0)}{2}
\sum_jq_j^2 ,
\end{eqnarray}
where the summation over $j$ is over all point charges $q_j$ in the
unit cell and $\tRi_{12}=\Ri+\rb_1-\rb_2$.

The recurrence relations associated with $\mOmega^{(n)}(R)$ and 
$\mOmega^{(n)}(0)$, respectively, take the form 
\begin{eqnarray}
\mOmega^{(n)}(R)=\frac{2\pi}{R}\int_0^\infty\sigma(r)r\,dr
\int_{|R-r|}^{R+r}\mOmega^{(n-1)}(y)y\,dy ,\label{Kq11}\\
\mOmega^{(n)}(0)=4\pi\int_0^\infty\sigma(r)\mOmega^{(n-1)}(r)r^2dr .
\label{Kq12}
\end{eqnarray}
The case of $n=1$ is then straightforward \cite{Kho108} and is described
by
\begin{eqnarray}
W^{(1)}(R)=4\pi\int_R^\infty\sigma(r)r\bigl(r-R\bigr)dr ,\label{Kq13}\\
\mOmega^{(1)}(0)=4\pi\int_0^\infty\sigma(r)r\,dr ,\label{Kq14}
\end{eqnarray}
providing that $\mOmega^{(0)}(R)=1/R$. In the case of $n=2$ one can obtain
\begin{eqnarray}
W^{(2)}(R)&=&4\pi^2\Bigl[\int_0^\infty dr_1\int_0^\infty dr_2A(r_1,r_2)
-\int_0^R dr_1\int_0^{R-r_1} dr_2\nonumber\\
&&{}\times A(r_1,r_2)-2\int_0^\infty dr_1\int_{R+r_1}^\infty dr_2B(r_1,r_2)
\Bigr] ,\label{Kq15}\\
\mOmega^{(2)}(0)&=&32\pi^2\int_0^\infty\sigma(r_1)r_1\,dr_1
\int_0^{r_1}\sigma(r_2)(r_2)^2\,dr_2 ,\label{Kq16}
\end{eqnarray}
where the following definitions
\begin{eqnarray}
A(r_1,r_2)=\sigma(r_1)\sigma(r_2)r_1r_2(R-r_1-r_2)^2 ,\label{Kq17}\\
B(r_1,r_2)=\sigma(r_1)\sigma(r_2)r_1r_2(R+r_1-r_2)^2 \label{Kq18}
\end{eqnarray}
are taken into account and notation (\ref{Kq9}) is also utilized.

Employed to the simple exponential spreading specified by
\begin{equation}\label{Kq19}
\sigma(r)=\frac{\alpha^3}{8\pi}\exp\bigl(-\alpha r\bigr) ,
\end{equation}
the latter results give rise to
\begin{eqnarray}
S(\hb)=\Bigl[1+\Bigl(\frac{2\pi|\hb|}{\alpha}\Bigr)^2\Bigr]^{-2} ,
\label{Kq20}\\
\tW^{(1)}(z)=\Bigl(1+\frac{z}{2}\Bigr)\exp(-z) ,\label{Kq21}\\
\tW^{(2)}(z)=\Bigl(1+\frac{11z}{16}+\frac{3z^2}{16}+\frac{z^3}{48}\Bigr)
\exp(-z) ,\label{Kq22}\\
\tW^{(3)}(z)=\Bigl(1+\frac{193z}{256}+\frac{65z^2}{256}+\frac{37z^3}{768}
+\frac{z^4}{192}+\frac{z^5}{3840}\Bigr)\exp(-z) ,\label{Kq23}\\
\mOmega^{(1)}(0)=\frac{\alpha}{2} ,\qquad\mOmega^{(2)}(0)=\frac{5\alpha}{16} ,
\qquad\mOmega^{(3)}(0)=\frac{63\alpha}{256} .\label{Kq24}
\end{eqnarray}
Here we go over to a new dimensionless variable $z=\alpha R$, so that the 
notations introduced above are modified as follows:
\begin{equation}\label{Kq25}
W^{(n)}(R)=\tW^{(n)}(z) .
\end{equation}

The other important case is associated with a Gaussian spreading function
\begin{equation}\label{Kq26}
\sigma_n(r)=\Bigl(\frac{n\mu^2}{\pi}\Bigr)^{3/2}\exp\bigl(-n\mu^2r^2\bigr) .
\end{equation}
The results appropriate to this case are of the form
\begin{eqnarray}
S_n^n(\hb)=\exp\Bigl(-\frac{\pi^2|\hb|^2}{\mu^2}\Bigr) ,\label{Kq27}\\
\tW^{(n)}(z)=\erfc(z)=\frac{2}{\sqrt{\pi}}\int_z^\infty\exp(-u^2)\,du ,
\label{Kq28}\\
\mOmega^{(n)}(0)=\frac{2\mu}{\sqrt{\pi}} .\label{Kq29}
\end{eqnarray}
The fact that the final results are independent of $n$ is the peculiar 
feature of the Ewald approach, as stressed earlier.

\section{Spatially confined spreading}\label{Sec3}
Let us now consider spherically symmetric spreading functions bounded by 
a radius $R_0$:
\begin{equation}\label{Lq1}
\sigma_k(r)=\cases{g_{k(s)}(r)& at $r\leq R_0$,\cr 0& at $r>R_0$,} 
\end{equation}
where we restrict ourselves to polynomials $g_{k(s)}$ of order $k$ which 
in turn result in $S_{k(s)}(\hb)\propto|\hb|^{-s}$. The meaning of the
parameter $s$ is defined therefrom. Then it is evident that 
relation (\ref{Kq9}) contributing to (\ref{Kq8}) still takes form 
(\ref{Kq25}), but at $z=R/R_0$. Furthermore, it will appear that 
$\tW_k^{(n)}(z)=0$ as $z\geq n$. This fact is natural for the Coulomb 
potential generated by a spherically symmetric charge distribution and 
implies that the sum over $i$ in (\ref{Kq8}) is actually finite and 
includes only unit cells nearest to a reference point.

There are different polynomials discussed in the literature \cite{Bert52,%
Heye81,Bert78,Jone56,Herz79,Ween75,Argy92,Kana55,Luty95}. Interested in 
principal particular cases of  spreading (\ref{Lq1}), we begin with a 
uniform spreading normalized by condition (\ref{Kq4}), as proposed by 
Bertaut \cite{Bert52}:
\begin{equation}\label{Lq2}
g_{0(2)}(r)=\frac{3}{4\pi R_0^3} .
\end{equation}
According to (\ref{Kq5}), relations (\ref{Lq1}) and (\ref{Lq2}) yield
\begin{equation}\label{Lq3}
S_{0(2)}(\hb)=\frac{3}{Y^2}\Bigl(\frac{\sin Y}{Y}-\cos Y\Bigr) ,
\end{equation}
where $Y=2\pi|\hb|R_0$ and so $s=2$ herein. If $n=1$, then the other 
quantities of interest, which are specified by the subscript $k(s)$ and
by the superscript $(n)$, are as follows:
\begin{eqnarray}
\tW_{0(2)}^{(1)}(z)=\cases{{\displaystyle\frac{(1-z)^2(2+z)}{2}}&at 
$0\leq z\leq1$,\cr 0&at $z>1$,}\label{Lq4}\\
\mOmega_{0(2)}^{(1)}(0)=\frac{3}{2R_0} .\label{Lq5}
\end{eqnarray}
The fact that $S_0(\hb)$ contains the factor $Y^{-2}$ enhances the rate
of convergence of a series over $\hb$ in (\ref{Kq8}) even at $n=1$.
This effect becomes stronger as $n$ increases. In the original approach
of Bertaut \cite{Bert52} the case of $n=2$ is considered. Making use
of (\ref{Kq15}) and (\ref{Kq16}), one can show that
\begin{eqnarray}
\tW_{0(2)}^{(2)}(z)=\cases{{\displaystyle\frac{(2-z)^4(10+8z+z^2)}{160}}&at 
$0\leq z\leq2$,\cr 0&at $z>2$,}\label{Lq6}\\
\mOmega_{0(2)}^{(2)}(0)=\frac{6}{5R_0} \label{Lq7}
\end{eqnarray}
in this case. The event of $n=3$ appears to be more complicated, though it
is still described by formulae (\ref{Kq11}) and (\ref{Kq12}) based on 
relations (\ref{Kq9}) and (\ref{Lq6}). The results can be cast in the form
\begin{eqnarray}
\mOmega_{0(2)}^{(3)}(0)=\frac{1269}{1280R_0} ,\label{Lq8}\\
\tW_{0(2)}^{(3)}(z)=\cases{M_1(z)+M_2(z) & at $0\leq z\leq1$ ,\cr
M_1(z) & at $1\leq z\leq3$,\cr 0 & at $z>3$,} \label{Lq9}
\end{eqnarray}
where
\begin{eqnarray}
M_1(z)=\frac{(3-z)^6(72+81z+18z^2+z^3)}{53760} ,
\label{Lq10}\\
M_2(z)=\frac{(1-z)^6(424+87z-6z^2-z^3)}{17920} .\label{Lq11}
\end{eqnarray}

It is important that, apart from enlarging the value of $n$, another way 
of enhancing the rate of convergence of the series over reciprocal vectors 
can be arrived at by increasing the order $k$ of $g_{k(s)}$ so as to 
enhance the degree $s$ of $Y$ in the denominator of $S_{k(s)}(\hb)$ 
\cite{Luty95}. In particular, one can show that
\begin{equation}\label{Lq12}
g_{1(3)}(r)=\frac{3(1-x)}{\pi R_0^3} ,
\end{equation}
where $x=r/R_0$, leads to
\begin{equation}\label{Lq13}
S_{1(3)}(\hb)=\frac{12}{Y^3}\Bigl(\frac{2(1-\cos Y)}{Y}-\sin Y\Bigr) .
\end{equation}
In this case
\begin{eqnarray}
\tW_{1(3)}^{(1)}(z)=\cases{(1-z)^3(1+z)& at $0\leq z\leq1$,\cr 0& at $z>1$,}
\label{Lq14}\\
\mOmega_{1(3)}^{(1)}(0)=\frac{2}{R_0} .\label{Lq15}
\end{eqnarray}
Extending the consideration to the case of $n=2$ here, we derive 
\begin{eqnarray}
\mOmega_{1(3)}^{(2)}(0)=\frac{52}{35R_0} ,\label{Lq16}\\
\tW_{1(3)}^{(2)}(z)=\cases{M_3(z)+M_4(z) & at $0\leq z\leq1$ ,\cr
M_3(z) & at $1\leq z\leq2$,\cr 0 & at $z>2$,} \label{Lq17}
\end{eqnarray}
where
\begin{eqnarray}
M_3(z)=\frac{(2-z)^6(2+4z+z^2)}{140} ,
\label{Lq18}\\
M_4(z)=\frac{(1-z)^7(3+z)}{35} .\label{Lq19}
\end{eqnarray}

Interested in polynomials of the lowest degree, the next several 
polynomials for the spreading function within this set are shown in 
\ref{App3}. Note that the couples of relations (\ref{Lq2}) and 
(\ref{Lq3}), (\ref{Lq12}) and (\ref{Lq13}) and finally (\ref{Xq1}) 
and (\ref{Xq2}) from \ref{App3} naturally agree with results
proposed in \cite{Heye81,Bert78,Jone56,Argy92,Kana55}. One can see 
that the set of spreading functions is reduced to terms of the form 
$(1-x)^k$ \cite{Heye81,Jenk79,Ween75} only at $k=1$ and $k=2$. This 
fact accounts for the known statement \cite{Jone56} that the simple 
form $(1-x)^k$ is not yet efficient at $k>2$.

Note that various polynomials are possible for a given $s$ if they are
not restricted to the lowest degree. Of different polynomials 
associated with $s=3$ and so competing with (\ref{Lq12}), here we 
consider the only one that is expected to be very effective \cite{Luty95} 
and is of the form
\begin{equation}\label{Lq20}
g_{2(3)}(r)=\frac{5x(1-x)}{\pi R_0^3} .
\end{equation}
Its Fourier transform is also known:
\begin{equation}\label{Lq21}
S_{2(3)}(\hb)=\frac{20}{Y^3}\Bigl(\frac{6\sin Y}{Y^2}-\frac{2+4\cos Y}{Y}
-\sin Y\Bigr) .
\end{equation}
All the other quantities appropriate to the case can be obtained in the manner
considered above. The case of $n=1$ is described by
\begin{eqnarray}
\tW_{2(3)}^{(1)}(z)=\cases{\frac{(1-z)^3(3+4z+3z^2)}{3}& at $0\leq z\leq1$,
\cr 0& at $z>1$,}\label{Lq22}\\
\mOmega_{2(3)}^{(1)}(0)=\frac{5}{3R_0} .\label{Lq23}
\end{eqnarray}
In the case of $n=2$, we reach 
\begin{eqnarray}
\mOmega_{2(3)}^{(2)}(0)=\frac{85}{63R_0} ,\label{Lq24}\\
\tW_{1(3)}^{(2)}(z)=\cases{M_5(z)-M_6(z) & at $0\leq z\leq1$ ,\cr
M_5(z) & at $1\leq z\leq2$,\cr 0 & at $z>2$,} \label{Lq25}
\end{eqnarray}
where
\begin{eqnarray}
M_5(z)=\frac{(2-z)^6(28+24z+42z^2+16z^3+3z^4)}{1512} ,
\label{Lq26}\\
M_6(z)=\frac{5(1-z)^7(7+4z+z^2)}{189} .\label{Lq27}
\end{eqnarray}

\section{Optimization of spreading parameters}
It is important that every spreading function is characterized by a
certain parameter. The problem of its optimization is traditional. 
In particular, the value of $\mu=2\sqrt{\pi}/d$ describing a Gaussian 
spreading function was used as a unique for the NaCl structure, where 
$d$ is the lattice spacing \cite{Ewal21,Dien48}. The situation 
associated with a simple exponential spreading appears to be quite 
similar \cite{Heye81,Birm58,Gool69}.

First of all, here we develop the treatment appropriate to this case.
As mentioned earlier \cite{Khol07}, it is based on the Coulomb 
characteristic of a Bravais lattice, the parameter put forward by Harris 
and Monkhorst \cite{Harr70}. In order to discuss this approach, we 
consider a Bravais lattice composed of unit point charges and immersed 
in a neutralizing uniform background. It is easy to show that the 
interaction of a background with the bulk potential field vanishes and 
the same is right for the background contribution to the sum over 
reciprocal lattice in expression (\ref{Kq10}) \cite{Khol07,Kho108,Harr70}. 
As a result, the effect of lattice summation in (\ref{Kq10}) is 
associated with the contribution of point charges alone. However, there 
is a remainder constituted of two simple finite terms there. One of those 
terms is determined by the last optional term in (\ref{Kq10}). The other 
one is the contribution of a background to the integral over real space 
and so is described by the negative of the quantity
\begin{equation}\label{Mq1}
G^{(n)}=\frac{8\pi^2n}{3v}\int_0^\infty r^4\sigma(r)\,dr ,
\end{equation}
keeping in mind that a background is of negative charge \cite{Kho108}.

In the case at hand equation (\ref{Kq2}) leads to $F(\hb)=1$. It implies 
that both the series over reciprocal and direct lattices in the original 
energy expressions are divergent and the spreading length, denoted in 
general as $\lambda$, actually forms cut-off parameters for both of those 
series in formula (\ref{Kq10}). Therefore it is expedient to cast 
the expression for the Coulomb characteristic $C$ of a Bravais lattice 
in the following schematic form:
\begin{equation}\label{Mq2}
\frac{A}{\lambda}+B\lambda^2=G^{(n)}_\lambda+\mOmega^{(n)}_\lambda(0)+C ,
\end{equation}
where the terms on the left-hand side stand for the contributions of
series over reciprocal and direct lattices, respectively, with singling out
their principal dependence on $\lambda$. Note that $C$ is the value of the
modulated lattice potential at the site of a unit point charge and so
$C/2=\Ec$, where $\Ec$ is the total energy per point charge in question
\cite{Khol07}.
It is significant that except for $C$, all the terms in relation (\ref{Mq2})
are positive. Moreover, according to definitions (\ref{Kq4}), (\ref{Kq12}) 
and (\ref{Mq1}), one can readily show that
\begin{equation}\label{Mq3}
G^{(n)}_\lambda\propto\lambda^2 ,\qquad
\mOmega^{(n)}_\lambda(0)\propto\lambda^{-1} .
\end{equation}
Therefore $G^{(n)}_\lambda$ and $\mOmega^{(n)}_\lambda(0)$ may be treated 
as counterparts to the direct-lattice series and to the reciprocal-lattice 
one in (\ref{Mq2}), respectively. Bearing in mind that $C$ is a constant, 
we draw a conclusion that an optimum relation between the terms on the 
left-hand side of formula (\ref{Mq2}) is associated with
a minimum of the right-hand side that is determined by the condition
\begin{equation}\label{Mq4}
\frac{d}{d\lambda}\Bigl[G^{(n)}_\lambda+\mOmega^{(n)}_\lambda(0)\Bigr]=0 .
\end{equation}

The Gaussian spreading (\ref{Kq26}) substituted into formula (\ref{Mq1}) 
at a given $n$ results in
\begin{equation}\label{Mq5}
G_\mu=\frac{\pi}{v\mu^2} .
\end{equation}
Note that this functional form is still independent of $n$. In this case 
$\lambda=1/\mu$. Inserting relations (\ref{Kq29}) and (\ref{Mq5}) into 
(\ref{Mq4}), we derive the result discussed earlier \cite{Khol04,Khol07}:
\begin{equation}\label{Mq6}
\mu^{\rm{opt}}=\sqrt{\pi}v^{-1/3} ,
\end{equation}
providing that the differentiation immediately with respect to $\mu$
in (\ref{Mq4}) is also admissible.

In the cases of simple exponential spreading discussed above, the
substitution of (\ref{Kq19}) into (\ref{Mq1}) yields
\begin{equation}\label{Mq7}
G^{(n)}_\alpha=\frac{8\pi n}{v\alpha^2} .
\end{equation}
Now we substitute issue (\ref{Mq7}) along with one of the relations described
by (\ref{Kq24}) into (\ref{Mq4}). The peculiar feature of relations
(\ref{Mq3}) implies that both the differentiation with respect to $\alpha$ 
and the differentiation with respect to $\lambda=1/\alpha$ in (\ref{Mq4})
also lead to the same result:
\begin{equation}\label{Mq8}
\alpha^{\rm{opt}}=\cases{{\displaystyle 2\Bigl(\frac{4\pi}{v}\Bigr)^{1/3}}&
at $n=1$ ,\cr
{\displaystyle 8\Bigl(\frac{\pi}{5v}\Bigr)^{1/3}}& at $n=2$ ,\cr
{\displaystyle 16\Bigl(\frac{\pi}{21v}\Bigr)^{1/3}}& at $n=3$ .}
\end{equation}
Note that the value of $\alpha^{\rm{opt}}$ increases when $n$ changes
from 1 to 3. It means that the initial spreading function becomes more
compact and thus the enhancing effect of spreading at $n=2$ and further 
at $n=3$ turns out to be somewhat restricted.  

Another case of optimization arises if the values of the spreading parameter 
$\alpha$, for definiteness, are supposed to depend on the cut-off parameter 
of summation $m$. If an integer value of $m$ is common to both summations 
over reciprocal and direct space, then the result for the energy can be 
written as $\Ec_{(n)}^{[m]}$. Apart from this quantity, it is also 
advantageous to consider the value of $\Ec_{(n)}^{[m+m_{\rm a}]}$, where 
$m_{\rm a}$ is a certain integer as well. The value of $\alpha_{(n)}^{[m]}$ 
associated with a given $m$ can be determined by the condition
\begin{equation}\label{Mq9}
\Ec_{(n)}^{[m+m_{\rm a}]}-\Ec_{(n)}^{[m]}=0 .
\end{equation} 
In order to understand this relation, one should point out that at least
two different situations arise here. Indeed, if $m_{\rm a}=1$, then 
formula (\ref{Mq9}) is apparently reduced to the condition of stability
with respect to small variations of $m$. The opposite limiting case
arises when $m_{\rm a}\gg1$. This case corresponds to the fact that
(\ref{Mq9}) evaluates all the rest removed within the cut-off procedure.
The requirement of zero value of this remainder is then natural. It
should be emphasized that both of these motifs are in accord with the 
statements discussed in the literature \cite{Humm95,Whee02}. In other
words, our latter treatment, albeit original, is developed in a quite
traditional manner, without recourse to any additional adjustable
function simulating the contribution of the lattice sum over direct
space \cite{Raja94}.

It is worth noting that according to (\ref{Mq9}), the value of the 
spreading parameter $\alpha_{(n)}^{[m]}$ is connected with the value of 
the cut-off parameter $m$. It is evident that this relation could be 
formally inverted, i.e., $m$ might be treated as a function of 
$\alpha_{(n)}^{[m]}$ as well. However, contrary to $\alpha_{(n)}$
that is continuous initially, $m$ is a discrete parameter and so a 
discrete set of values $\alpha_{(n)}^{[m]}$ appropriate to different 
$m$ actually arises as a solution of (\ref{Mq9}).
 
One more possibility to optimize the calculation of lattice series
is associated with the fact that the lattice sum in direct space 
converges faster then that in reciprocal space. Therefore it may be
advantageous to choose the cut-off parameter 
$m_{\scriptscriptstyle\rm R}$ truncating the direct lattice sum larger 
than the cut-off parameter $m$ applied to the corresponding reciprocal 
lattice sum. In this case the correlation between the cut-off parameter 
$m_{\scriptscriptstyle\rm R}$ and the spreading parameter $\alpha_{(n)}$ 
may be of interest, providing that the third parameter $m$ is fixed 
\cite{Raja94,Humm95,Whee02}. The foregoing treatment based on (\ref{Mq9}) 
can be easily extended to the present case if we adopt
\begin{equation}\label{Mq10}
\Ec_{(n)}^{[m+m_{\rm a},\; m_{\scriptscriptstyle\rm R}+m_{\rm a}]}
-\Ec_{(n)}^{[m,\; m_{\scriptscriptstyle\rm R}]}=0 ,
\end{equation}
where we conjecture that the increment $m_{\rm a}$ in both $m$ and
$m_{\scriptscriptstyle\rm R}$ is the same, for simplicity. Note that the
two limitimg cases mentioned above as associated with $m_{\rm a}$ still 
take place. As far as the choice of $m_{\scriptscriptstyle\rm R}$ is
concerned, it turns out that this value is to be as large as possible.
Actually its value is restricted by machine accuracy. On the other hand,
its effect on $\alpha_{(n)}$ is strong, as will be shown later on.

In the events of confined spreading functions restricted by $R_0$, the 
Bertaut version dealing with non-overlapping charge distributions is the 
most popular. As is well known, the sum over direct space is then zero and 
so it does not contribute to the result that is determined by the sum over 
reciprocal space alone \cite{Bert52,Bert78}. Keeping in mind that the value 
$R_0$ must still be as large as possible, one may conclude that $R_0$ is 
to be half the nearest interatomic distance \cite{Bert52,Heye81,Jone56,%
Herz79,Ween75,Argy92,John61,Jenk75,Herz81}. This convention is sustained in 
modern papers devoted to this subject as well \cite{Luty95}. As a result, 
the assessment of the accuracy of computation is reduced to the classical 
consideration of the cut-off effect upon summation over reciprocal space 
\cite{Temp55,Jenk79}.

However, it is expedient to note that the aforementioned restriction on
$R_0$ may be convenient, but is not principal. Indeed, some further growth
of $R_0$ results in the appearance of a certain direct sum that is finite
and can be calculated rigorously. On the other hand, the larger $R_0$ the
faster convergence of the series over reciprocal space. This is the reason 
to make use of the largest value of $R_0$ consistent with machine accuracy 
of evaluation of the direct sum \cite{Kawa99,Herz79,Luty95}. The latter is 
the problem typical of practical summation.

\section{Effect of infinite normalized spreading functions}
Here we apply the foregoing results to the classical NaCl structure
composed of point charges, which remains attractive as a model system 
\cite{Argy92,Herz81,Tyag04}. As known, a face-centred cubic Bravais 
lattice describes this structure, with two sites per unit cell. If $d$ 
is the edge of an elementary cube, then $v=d^3/4$ and the basis vectors 
for point charges $\pm q$ may be chosen as:
\begin{equation}\label{Wq1}
+q:\;\bb_1=(0,0,0),\qquad -q:\;\bb_2=(d/2,0,0) .
\end{equation}
In terms of the elementary translations
\begin{equation}\label{Wq2}
\ab_1=\frac{d(1,1,0)}{2} ,\qquad \ab_2=\frac{d(1,0,1)}{2},\qquad
\ab_3=\frac{d(0,1,1)}{2} ,
\end{equation}
an arbitrary translation vector is of the form
\begin{equation}\label{Wq3}
\Ri=m_1\ab_1+m_2\ab_2+m_3\ab_3 ,
\end{equation}
where $m_j$ are integers. The elementary reciprocal lattice translations
appropriate to the vectors in (\ref{Wq2}) are defined by the scalar product
$(\ab_i\hb_j)=\delta_{ij}$, where $\delta_{ij}$ is the Kronecker delta, and 
are as follows:
\begin{equation}\label{Wq4}
\hb_1=\frac{(1,1,-1)}{d} ,\qquad \hb_2=\frac{(1,-1,1)}{d},\qquad
\hb_3=\frac{(-1,1,1)}{d} .
\end{equation}
They compose a general reciprocal lattice vector of the form 
\begin{equation}\label{Wq5}
\hb=m_1\hb_1+m_2\hb_2+m_3\hb_3 .
\end{equation}
Note that definitions (\ref{Wq2}) and (\ref{Wq4}) are conventional 
\cite{Kitt76}.

On taking formulae (\ref{Kq2}), (\ref{Kq3}) and (\ref{Wq1}) into account,
expression (\ref{Kq10}) for the energy of interest is transformed into
\begin{eqnarray}\label{Wq6}
\fl\Ec_{(n)}&=&\frac{q^2}{\pi v}\sump_\hb\frac{\{1-\cos[2\pi
(\hb\bb_2)]\}S^n(\hb)}{|\hb|^2}+q^2\sump_i\Bigl[\frac{W^{(n)}
(|\Ri|)}{|\Ri|}-\frac{W^{(n)}(|\Ri+\bb_2|)}{|\Ri+\bb_2|}\Bigr]\nonumber\\
\fl&&{}-q^2\mOmega^{(n)}(0) .
\end{eqnarray}

According to (\ref{Wq2})--(\ref{Wq5}), the parameters necessary for the 
further summation take the form
\begin{eqnarray}
\fl|\Ri|=\frac{d}{2}\Bigl[(m_1+m_2)^2+(m_2+m_3)^2+(m_3+m_1)^2\Bigr]^{1/2} ,
\label{Wq7}\\
\fl|\Ri+\bb_2|=\frac{d}{2}\Bigl[(m_1+m_2+1)^2+(m_2+m_3)^2
+(m_3+m_1)^2\Bigr]^{1/2} ,\label{Wq8}\\
\fl|\hb|=\frac{1}{d}\Bigl[(m_1+m_2-m_3)^2+(m_2+m_3-m_1)^2
+(m_3+m_1-m_2)^2\Bigr]^{1/2} ,\label{Wq9}\\
\fl 1-\cos[2\pi(\hb\bb_2)]=\cases{0&at $m_1+m_2-m_3$ even,\cr
2&at $m_1+m_2-m_3$ odd.}\label{Wq10}
\end{eqnarray}
Note that the latter relation is very specific \cite{Argy92}. Employing 
relations (\ref{Wq7})--(\ref{Wq10}) in formula (\ref{Wq6}), we propose 
\begin{table}[t]
\caption{The Coulomb energy $\Ec$, in units of $q^2/d$, per unit cell 
of the NaCl point-charge lattice. The calculation is based on a 
Gaussian charge spreading with $\mu=2\sqrt{\pi}/d$ proposed originally 
by Ewald ($\Ec_{\rm Ew}$) and with $\mu=\sqrt{\pi}v^{-1/3}$ in agreement 
with our proposal ($\Ec_{\rm our}$). The results, with significant 
figures only, are shown in dependence on $m$ restricting actual ranges 
of summation over direct and reciprocal space, in accord with condition 
(\ref{Wq11}).}\label{Table1}
\begin{indented}
\item[]\begin{tabular}{@{}rclcl}
\br
m&$\;\;$&$\hspace{2em}\Ec_{\rm Ew}$&$\;\;$&$\hspace{3em}\Ec_{\rm our}$\\
\mr
$1$&&$-3.5              $&&$-3.5              $\\
$2$&&$-3.49513          $&&$-3.4951292        $\\
$3$&&$-3.49512918927    $&&$-3.49512918926636{\lefteqn{^{\rm a}}} $\\
\br
\end{tabular}
\item[]{$^{\rm a}\,$This value agrees with the result of Sakamoto \cite{Saka58}
.}
\end{indented}
\end{table}
that the parameters of summation are restricted by a common condition
\begin{equation}\label{Wq11}
|m_j|\leq m ,
\end{equation}
where an integer $m$ is varying. The rate of convergence of series in 
equation (\ref{Wq6}) will be then studied in dependence on $m$.

Here we start from a Gaussian spreading function that is a classical example
\cite{Kawa99,Khol07,Heye81}. Bearing in mind that the effect of multiple 
charge spreading is reduced only to some definite scaling of Gaussian 
parameters in relations (\ref{Kq27})--(\ref{Kq29}), we consider the case of 
a unique optimal value of a Gaussian parameter described by (\ref{Mq6}). The 
value of the Madelung energy is shown in table \ref{Table1} in dependence on 
$m$. The effect of the original value of $\mu=2\sqrt{\pi}/d$ proposed by 
Ewald \cite{Ewal21,Dien48} is demonstrated in the same table for comparison. 
Table 1 shows that the rate of convergence is actually fantastic in both 
these cases, though our choice appears to be somewhat more advanced. Such 
results are in general expected \cite{Jenk79}. This is the reason that we 
will not discuss further possibilities of optimization addressed to a 
Gaussian spreading function \cite{Humm95}. 

At this stage it is worth noting that the accuracy of results specified by
Gaussian spreading function drops drastically if we restrict ourselves to 
spherical domains of summation over both the reciprocal and direct lattices. 
This fact is known \cite{Raja94}. Unfortunately, such a loss in accuracy 
caused by spherical modes of summation in the series at hand turns out to be 
typical of all the other cases discussed below. It means that modes of 
summation supporting the crystal symmetry are more advantageous for perfect 
crystals. On the other hand, very popular schemes of spherical summation seem 
to be essential in applications describing disordered systems 
\cite{Humm95,Whee02}.

In the cases of a simple exponential spreading formulae
(\ref{Kq20})--(\ref{Kq24}) and (\ref{Mq8}) are utilized in (\ref{Wq6}) first of
all. The results of summation are listed in table \ref{Table2}. As anticipated,
\begin{table}[t]
\caption{The Coulomb energy $\Ec$, in units of $q^2/d$, per unit cell of 
the NaCl point-charge lattice. The calculation is based on a simple 
exponential charge spreading with optimal values of $\alpha_{(n)}$ described
by relation (\ref{Mq7}) for $n=1$, $n=2$ and $n=3$. The results are shown in 
dependence on $m$ in condition (\ref{Wq11}) again.}\label{Table2}
\begin{indented}
\item[]\begin{tabular}{@{}rclclcl}
\br
m&$\;\;$&$\hspace{1.5em}\Ec_{(1)}$&$\;\;$&$\hspace{2.3em}\Ec_{(2)}$&$\;\;$&$\hspace{3em}\Ec_{(3)}$\\
\mr
$ 5 $&&$-3.5   $&&$-3.4951      $&&$-3.495129        $\\
$10 $&&$-3.5   $&&$-3.495129    $&&$-3.495129189     $\\
$15 $&&$-3.495 $&&$-3.4951292   $&&$-3.49512918927   $\\
$20 $&&$-3.495 $&&$-3.49512919  $&&$-3.4951291892664 $\\ 
$25 $&&$-3.495 $&&$-3.495129189 $&&$-3.4951291892664 $\\ 
$30 $&&$-3.4951$&&$-3.4951291893$&&$-3.4951291892664 $\\ 
$35 $&&$-3.4951$&&$-3.4951291893$&&$-3.49512918926636{\lefteqn{^{\rm a}}}$\\
\br
\end{tabular}
\item[]{$^{\rm a}\,$This value agrees with the result of Sakamoto \cite{Saka58}
.}
\end{indented}
\end{table}
the tendency towards enhancing the rate of convergence takes place upon
increasing the order of multiplicity $n$. This effect is quite general and
has been pointed out by Bertaut \cite{Bert78} in connection with the particular
cases of $n=1$ and $n=2$ at spatially confined spreading functions, as will 
be discussed in the sequel. 

However, table \ref{Table2} also shows that, contrary to a Gaussian spreading 
function, the application of a simple exponential spreading with fixed optimal 
values of $\alpha_{(n)}$ results in much more moderate rate of convergence of 
lattice series in question. Therefore further efforts towards optimizing lattice 
calculations seem to be necessary here. The next step of optimization discussed
is associated with a possible variation of $\alpha_{(n)}$ in dependence on $m$.
With the help of condition (\ref{Mq9}), this effect can be readily taken into 
account. Here two particular cases of $m_{\rm a}=1$ and $m_{\rm a}=10$ are 
studied. The results of calculation are compiled in table \ref{Table3}. We see 
\begin{table}[b]
\caption{The Coulomb energy $\Ec$, in units of $q^2/d$, per unit cell of the NaCl 
point-charge lattice in the case similar to that in table \ref{Table2}, but with
$\alpha_{(n)}$ of which variation with $m$ is specified by (\ref{Mq9}) at
$m_{\rm a}=1$ and $m_{\rm a}=10$, respectively.  
}\label{Table3}
\parindent=3.9em
\footnotesize{
\begin{tabular}{@{}rllrlrl}
\br
$m$&$\hspace{0.6em}\alpha_{(1)}$&$\hspace{1.5em}\Ec_{(1)}$&$\alpha_{(2)}\hspace{0.6em}$&$\hspace{2.5em}\Ec_{(2)}$&$\alpha_{(3)}\hspace{0.6em}$&$\hspace{3em}\Ec_{(3)}$\\
\mr
\multicolumn{7}{c}{case of $m_{\rm a}=1$}\\
\mr
$ 1$&$6.8265$&$-3.5    $&$11.2028$&$-3.5          $&$14.4677$&$-3.495           $\\
$ 3$&$5.0228$&$-3.5    $&$ 8.7002$&$-3.49513      $&$11.5983$&$-3.4951292       $\\
$ 5$&$4.0395$&$-3.495  $&$ 7.2189$&$-3.4951292    $&$ 9.8319$&$-3.49512919      $\\
$ 7$&$3.4073$&$-3.495  $&$ 6.2024$&$-3.49512919   $&$ 8.5572$&$-3.49512918927   $\\
$ 9$&$2.9639$&$-3.49513$&$ 5.4620$&$-3.49512919   $&$ 7.6008$&$-3.495129189266  $\\
$11$&$2.6336$&$-3.49513$&$ 4.8970$&$-3.4951291893 $&$ 6.8568$&$-3.4951291892664 $\\
$13$&$2.3769$&$-3.49513$&$ 4.4502$&$-3.49512918927$&$ 6.2618$&$-3.49512918926636
{\lefteqn{^{\rm a}}}$\\
\mr
\multicolumn{7}{c}{case of $m_{\rm a}=10$}\\
\mr
$ 1$&$6.5898$&$-3.495  $&$11.1262$&$-3.4951292   $&$14.4308$&$-3.4951291893    $\\
$ 2$&$5.5308$&$-3.495  $&$ 9.6293$&$-3.4951292   $&$12.6762$&$-3.49512918927   $\\
$ 3$&$4.8102$&$-3.495  $&$ 8.6223$&$-3.49512919  $&$11.5650$&$-3.49512918927   $\\
$ 4$&$4.2661$&$-3.4951 $&$ 7.7974$&$-3.49512919  $&$10.5979$&$-3.49512918927   $\\
$ 5$&$3.8453$&$-3.4951 $&$ 7.1272$&$-3.495129189 $&$ 9.7854$&$-3.4951291892664 $\\
$ 6$&$3.5093$&$-3.49513$&$ 6.5721$&$-3.4951291893$&$ 9.0943$&$-3.4951291892664 $\\
$ 7$&$3.2346$&$-3.49513$&$ 6.1058$&$-3.4951291893$&$ 8.5019$&$-3.4951291892664 $\\
$ 8$&$3.0052$&$-3.49513$&$ 5.7084$&$-3.4951291893$&$ 7.9888$&$-3.49512918926636
{\lefteqn{^{\rm a}}}$\\
\br
\end{tabular}
{\par$^{\rm a}\,$This value agrees with the result of Sakamoto \cite{Saka58}.}
}
\end{table}
that the tendency towards reducing the value of $\alpha_{(n)}$ upon growing 
\begin{table}[t]
\caption{The Coulomb energy $\Ec$, in units of $q^2/d$, per unit cell of 
the NaCl point-charge lattice in the case of a simple exponential charge 
spreading like that in table \ref{Table3}, but with $m=1$ and 
$m_{\scriptscriptstyle\rm R}=22$ restricting lattice summations over 
reciprocal and direct space, respectively, in accord with (\ref{Wq11}). 
Then the values of $\alpha_{(n)}$ at $n=1$, 2 and 3 are determined by 
(\ref{Mq10}) at $m_{\rm a}=1$ and $m_{\rm a}=10$.}\label{Table4}
\begin{indented}
\item[]\begin{tabular}{@{}rclclcl}
\br
$m_{\rm a}$&&$n$&&$\hspace{0.5em}\alpha_{(n)}$&&$\hspace{3em}
\Ec_{(n)}$\\
\mr
$1$&&$1$&&$1.0346$&&$-3.4951$\\
   &&$2$&&$1.9312$&&$-3.4951292$\\
   &&$3$&&$2.6808$&&$-3.495129189$\\
\mr   
$10$&&$1$&&$1.0681$&&$-3.49512919$\\
    &&$2$&&$1.9562$&&$-3.49512918927$\\
    &&$3$&&$2.6986$&&$-3.495129189266364{\lefteqn{^{\rm a}}}$\\
\br
\end{tabular}
\item[]{$^{\rm a}\,$This value agrees with the result of Sakamoto \cite{Saka58}.}
\end{indented}
\end{table}
$m$ is common to all the events of $n$ under consideration. It implies 
that the effective charge distributions become more diffuse as the range 
of summation increases. On the other hand, the multiple spreading leads 
to larger values of $\alpha_{(n)}$. This output agrees with the results of 
(\ref{Mq8}). It is important that the enhancement of the rate of 
convergence arises in both the cases of $m_{\rm a}$, but the effect at 
$m_{\rm a}=10$ is somewhat stronger than that at $m_{\rm a}=1$. 

The latter trend can be fruitful when we go over to the next step of
optimization based on relation (\ref{Mq10}). The event of $m=1$ and
$m_{\scriptscriptstyle\rm R}=22$ is studied, providing that the values
$m_{\rm a}=1$ and $m_{\rm a}=10$ are utilized in (\ref{Mq10}). The 
corresponding results are listed in table \ref{Table4}. Table 
\ref{Table4} shows that the values of $\alpha_{(n)}$ corresponding to 
each value of $n$ appear to be close, but the effect of $m_{\rm a}=10$ is 
more pronounced again. 

\section{Trends in the application of confined spreading functions}
As far as confined spreading functions are concerned, here we restrict 
ourselves to the situation opposite to the classical one described by 
charge distributions non-overlapping after spreading \cite{Bert52,Bert78}. 
In other words, we propose that the parameter $R_0$ in (\ref{Lq1}) is greater 
than the lattice parameter $d$. It is important that the sum over direct space 
is always finite in such a case and so it can be counted with a high precision. 
In practice our choice of $R_0$ is actually restricted by machine accuracy. 
As a result, optimum values of $R_0$ appear to be different and depend 
on each particular case under consideration. For example, if $k=0$, then we 
adopt $R_0=7d$ upon considering all three cases for $n$ from one to three. 
Note that the value of $m$ in (\ref{Wq11}) restricts only the reciprocal 
lattice series now. The computation based on formulae (\ref{Lq2})--(\ref{Lq11}) 
substituted into (\ref{Wq6}) gives rise to the results presented in table 
\ref{Table5}. Table \ref{Table5} shows 
\begin{table}[b]
\caption{The specific Coulomb energy $\Ec$, in units of $q^2/d$, for the NaCl
point-charge lattice is obtained in dependence on the restricting parameter $m$
at a fixed value of $R_0=7d$ common to all the cases of the confined polynomial 
spreading function $g_{0(2)}(r)$ defined by formula (\ref{Lq2}). The cases of
$\Ec_{(n)}$ at $n=1$, $n=2$ and $n=3$, which are specified by 
(\ref{Lq3})--(\ref{Lq11}) in formula (\ref{Wq6}) are considered.}\label{Table5}
\begin{indented}
\item[]\begin{tabular}{@{}rclclcl}
\br
$m$&&$\hspace{1.5em}\Ec_{(1)}$&&$\hspace{2em}\Ec_{(2)}$&&$\hspace{3em}\Ec_{(3)}$\\
\mr
$ 0$&&$-3.5     $&&$-3.49513     $&&$-3.495129189       $\\
$ 1$&&$-3.495   $&&$-3.4951292   $&&$-3.49512918927     $\\
$ 2$&&$-3.495   $&&$-3.4951292   $&&$-3.49512918927     $\\
$ 3$&&$-3.495   $&&$-3.4951292   $&&$-3.4951291892664   $\\
$ 4$&&$-3.495   $&&$-3.49512919  $&&$-3.4951291892664   $\\
$ 5$&&$-3.495   $&&$-3.49512919  $&&$-3.4951291892664   $\\
$10$&&$-3.4951  $&&$-3.49512919  $&&$-3.495129189266364 $\\
$15$&&$-3.495129$&&$-3.4951291893$&&$-3.4951291892663644{\lefteqn{^{\rm a}}}$\\
\br
\end{tabular}
\item[]{$^{\rm a}\,$This value agrees with the result of Sakamoto \cite{Saka58}
(see also references \cite{Hajj72,CraD87}).}
\end{indented}
\end{table}
that the accuracy of the output obtained in these cases enhances rapidly upon 
changing $n$ from 1 to 3. This result agrees with the general tendency mentioned
above \cite{Bert78}. By the way, it is curious that the value at the bottom of the 
first column appears to be in between the values determined by $m=0$ and $m=1$ in 
the second column and the same is right for the second and third columns as well. 
Of course, this fact seems to be occasional, but it is expressive. 

Note that the case of $m=0$ here and below describes the contribution of 
the sum over direct space alone. The corresponding results shown in table 
\ref{Table5} give evidence that the summation over real space alone can be 
very precise and so may be treated as one more efficient approach to the direct 
lattice summation problem \cite{Khol04,Heye81,Khol06}. Within such a treatment 
the series over reciprocal lattice vectors may in turn be regarded as an 
additional correcting contribution to the direct-space one. It is natural that 
it leads to the further essential enhancement of the overall accuracy.

At the next step we consider the same energy counted with making use of the 
spreading function $g_{1(3)}(r)$ specified by (\ref{Lq12}) and (\ref{Lq13}) at
$R_0=6.5d$ as an optimal value. The corresponding cases of $n=1$ and $n=2$ are
determined by formulae (\ref{Lq14})--(\ref{Lq15}) and (\ref{Lq16})--(\ref{Lq19}), 
respectively. We also examine the case of $g_{2(3)}(r)$ that is optimized at 
$R_0=7d$ and is determined by relations (\ref{Lq20})--(\ref{Lq27}), with including
the events of $n=1$ and $n=2$ as well. The results of calculation are given in
table \ref{Table6}. On comparing these results with the first two columns of
\begin{table}[t]
\caption{The specific Coulomb energy $\Ec$, in units of $q^2/d$, for the NaCl
point-charge lattice obtained with making use of either $g_{1(3)}(r)$ defined
by (\ref{Lq12}) at $R_0=6.5d$ or $g_{2(3)}(r)$ defined by (\ref{Lq20}) at 
$R_0=7d$, in dependence on the cut-off parameter $m$. The cases of $n=1$ 
and $n=2$ for both these spreading functions are considered together for 
comparison.}\label{Table6}
\parindent=1.7em
\footnotesize{
\begin{tabular}{@{}rclclclcl}
\br
&$\;$&\multicolumn{3}{c}{case of $g_{1(3)}(r)$}&$\;$&
\multicolumn{3}{c}{case of $g_{2(3)}(r)$}\\
\cline{3-5}\cline{7-9}\\[-6pt]
m&&$\hspace{1.8em}\Ec_{(1)}$&&$\hspace{3em}\Ec_{(2)}$&&$\hspace{1.8em}\Ec_{(1)}$
&&$\hspace{3em}\Ec_{(2)}$\\
\mr
$ 0$&&$-3.495     $&&$-3.4951292         $&&$-3.495      $&&$-3.4951292       $\\
$ 1$&&$-3.49513   $&&$-3.4951291893      $&&$-3.4951     $&&$-3.495129189     $\\
$ 2$&&$-3.49513   $&&$-3.4951291893      $&&$-3.49513    $&&$-3.49512918927   $\\
$ 3$&&$-3.495129  $&&$-3.49512918927     $&&$-3.49513    $&&$-3.49512918927   $\\
$ 4$&&$-3.495129  $&&$-3.49512918927     $&&$-3.49513    $&&$-3.49512918927   $\\
$ 5$&&$-3.495129  $&&$-3.49512918927     $&&$-3.49513    $&&$-3.49512918927   $\\
$10$&&$-3.495129  $&&$-3.4951291892664   $&&$-3.495129   $&&$-3.4951291892664 $\\
$15$&&$-3.4951292 $&&$-3.4951291892664   $&&$-3.495129   $&&$-3.4951291892664 $\\
$20$&&$-3.49512919$&&$-3.49512918926636{\lefteqn{^{\rm a}}}  $&&$-3.49512919 $&&$-3.49512918926636{\lefteqn{^{\rm a}}}$\\
\br
\end{tabular}
{\par$^{\rm a}\,$This value agrees with the result of Sakamoto \cite{Saka58}}
}
\end{table}
table \ref{Table5}, it is evident that the latter issues are more efficient in
the case of $n=1$ and the same is right for $n=2$. The tendency that every case
of $n=2$ is much more accurate than the corresponding case of $n=1$ is also 
maintained. On the other hand, table \ref{Table6} shows that any advantage of 
$g_{2(3)}(r)$ over $g_{1(3)}(r)$, that is expected for non-overlapping spreading
functions \cite{Luty95}, is not realized in the cases of large $R_0$. Moreover, the case 
of $g_{1(3)}(r)$ at $n=2$ appears to be somewhat more precise, keeping in mind 
that this effect is achieved at a less value of $R_0$.

It is significant that the enhancement of accuracy takes place along the set of 
spreading functions considered in \ref{App3} even at $n=1$. Note that the formulae
associated especially with $g_{4(5)}(r)$ and $g_{5(6)}(r)$ are rather novel. This 
is the reason that the results appropriate to those cases and obtained at the 
corresponding optimum value of $R_0=15d$ are listed in table \ref{Table7}. Indeed, 
\begin{table}
\caption{The specific Coulomb energy $\Ec$, in units of $q^2/d$, for the NaCl
point-charge lattice is obtained in dependence on the restricting parameter $m$
at a fixed value of $R_0=15d$ common to all the cases of the confined polynomials 
specified by (\ref{Xq1})--(\ref{Xq12}) in \ref{App3} at $n=1$.}\label{Table7}
\begin{indented}
\item[]\begin{tabular}{@{}rclclcl}
\br
m&&\multicolumn{5}{c}{$\Ec_{(1)}$}\\
\cline{3-7}\\[-6pt]
& &case of $g_{2(4)}(r)$& &case of $g_{4(5)}(r)$& &case of $g_{5(6)}(r)$\\
\mr
$ 0$&&$-3.49513     $&&$-3.49512919      $&&$-3.49512919$\\
$ 1$&&$-3.495129    $&&$-3.49512919      $&&$-3.495129189$\\
$ 2$&&$-3.4951292   $&&$-3.4951291893    $&&$-3.4951291893$\\
$ 3$&&$-3.4951292   $&&$-3.4951291893    $&&$-3.49512918927$\\
$ 4$&&$-3.49512919  $&&$-3.49512918927   $&&$-3.49512918927$\\
$ 5$&&$-3.49512919  $&&$-3.49512918927   $&&$-3.49512918927$\\
$10$&&$-3.49512919  $&&$-3.4951291892664 $&&$-3.4951291892664$\\
$15$&&$-3.495129189 $&&$-3.49512918926636$&&$-3.4951291892664$\\
$20$&&$-3.495129189 $&&$-3.49512918926636$&&$-3.49512918926636$\\
$25$&&$-3.495129189 $&&$-3.49512918926636$&&$-3.49512918926636$\\
$30$&&$-3.495129189 $&&$-3.49512918926636$&&$-3.495129189266364$\\
$35$&&$-3.4951291893$&&$-3.49512918926636$
&&$-3.4951291892663644{\lefteqn{^{\rm a}}}$\\ 
\br
\end{tabular}
\item[]{$^{\rm a}\,$This value agrees with the result of Sakamoto \cite{Saka58}
(see also references \cite{Hajj72,CraD87}).}
\end{indented}
\end{table}
the rate of convergence increases as $k$ grows. Of course, this effect turns out 
to be less prominent in comparison with the cases of $n=2$ and $n=3$ mentioned 
above. Nevertheless, at $k=5$ the limiting precision adopted in our calculations 
is eventually attained at as well. 

Note that if $n$ is odd, then the contribution of the reciprocal lattice sum 
tends to the exact value in an oscillatory manner due to trigonometric 
functions describing $S_{k(s)}(\hb)$ appropriate to all the events in this
section \cite{Heye81,Argy92}.

Some final comment is necessary on the question why all the foregoing tables 
contain limiting energy values with a rather large number of significant figures. 
Indeed, one may think that the lattice parameters of real structures are usually 
known up to four, at best six, figures only \cite{Wyck64}. Nevertheless,
even in this case the numerical calculation should be a bit more accurate. 
However, the reason of our ultimate accuracy is quite different. As shown, the 
analytic accuracy of a series is associated with the number of unit cells taken 
into account and restricted by a cut-off parameter introduced upon series 
computation. On the other hand, restricted by machine accuracy, the overall
accuracy of computation depends on the total number of implemented operations 
and so is connected anyhow with the total number of point charges taken into 
account. The interplay between these tendencies implies that the effect of 
machine accuracy may be predominant if the number of charges per unit cell 
increases. In other words, an ultimate accuracy achieved for model structure 
with a small unit cell may be regarded as a guaranty of a sufficient accuracy 
while each of the approaches developed above is applied to modern compounds 
with large unit cells. 

\section{Conclusion}
In summary, it has been confirmed that the effect of a multiple charge spreading 
results in increasing the rate of convergence of the lattice series. In other
words, we deal with one more way to make the convergence faster, providing
that this effect becomes stronger and stronger with every further repetition of 
the procedure of spreading. It is important that this is a property common to 
both the classes of infinite and confined normalized spreading functions. 

The optimization of the shape of spreading is more traditional in the problem
of enhancing the rate of convergence. Nevertheless, we have proposed some
novel approaches to this task. In particular, for infinite normalized functions
of spreading two different situations are considered. In the case of a fixed
spreading parameter, the universal approach of optimization is based on the 
investigation of the concomitant terms contributing to the Coulomb energy, 
which are independent of the lattice sums. The problem of optimizing the 
spreading parameter in dependence on the cut-off parameters of summation 
has been solved with the help of some conditions imposed on the remainder of
the Coulomb energy, providing that this remainder is also treated in a truncated
form. In the case of confined spreading functions the optimization in question
is reduced to separating any main set of polynomials ensuring the progressive
enhancement of the rate of convergence of the sum over reciprocal space.
 
It is found that the effect of optimization becomes much more efficient if 
the multiple spreading is applied as well. The only case independent of the 
multiple spreading is described by a Gaussian spreading function due to its 
invariance with respect to spreading in a multiple manner \cite{Kho108}. 
Moreover, it turns out that the convergence with a Gaussian spreading function 
is the most prominent even at a fixed optimized spreading parameter. This fact 
gives evidence that a Gaussian spreading seems to be the most suitable one in 
the treatment based on the charge spreading as a whole.

On the other hand, the investigation of confined functions of spreading shows 
that it is fruitful to choose larger values of the parameter restricting 
those functions. Of course, a certain finite sum over sites of the direct 
lattice then appears. However, there is no problem with its calculation.
On the other hand, the contribution of the sum over reciprocal lattice 
becomes smaller, up to the case where the latter sum describes only a
small correction to the result of direct summation. Note that this trend is
eventually common to all the examples discussed above. By the way, it implies 
one more approach to the problem of direct summation of Coulomb series in 
crystals.

\appendix
\section{Particular forms of a polynomial spreading function}\label{App3}
Here we continue the fundamental series of polynomials beginning with
(\ref{Lq2}) and (\ref{Lq12}), which produce $S(\hb)$ with the enhancing
power of $Y$ in the denominator, providing that the polynomial degree be
a minimum. Note that the case of $n=1$ alone will be considered here.
So, the next example of such a polynomial, with including the concomitant
relations, is of the form 
\begin{eqnarray}
g_{2(4)}(r)=\frac{15(1-x)^2}{2\pi R_0^3} ,\label{Xq1}\\
S_{2(4)}(\hb)=\frac{60}{Y^4}\Bigl(2+\cos Y-\frac{3\sin Y}{Y}\Bigr) ,
\label{Xq2}\\
W_{2(4)}^{(1)}(z)=\cases{{\displaystyle\frac{(1-z)^4(2+3z)}{2}}
\hspace{-0.2em}&at $0\leq z\leq1$,\cr 0&at $z>1$,}\label{Xq3}\\
\mOmega_{2(4)}^{(1)}(0)=\frac{5}{2R_0} ,\label{Xq4}
\end{eqnarray}
where the notations of section \ref{Sec3} are utilized. 

The next polynomial of spreading within the set under consideration
is not simply defined by $(1-x)^3$, but it is described as
\begin{equation}\label{Xq5}
g_{4(5)}(r)=\frac{105(1-x)^3(1+3x)}{16\pi R_0^3} ,
\end{equation}
with the concomitant quantities
\begin{eqnarray}
S_{4(5)}(\hb)=\frac{630}{Y^5}\Bigl(\sin Y+\frac{8+7\cos Y}{Y}
-\frac{15\sin Y}{Y^2}\Bigr) ,\label{Xq6}\\
W_{4(5)}^{(1)}(z)=\cases{{\displaystyle\frac{(1-z)^5(8+19z+15z^2)}{8}}
\hspace{-0.2em}&at $0\leq z\leq1$,\cr 0&at $z>1$,}\label{Xq7}\\
\mOmega_{4(5)}^{(1)}(0)=\frac{21}{8R_0} .\label{Xq8}
\end{eqnarray}
The relations specifying one more pattern of this set are of the form 
\begin{eqnarray}
g_{5(6)}(r)=\frac{21(1-x)^4(1+4x)}{2\pi R_0^3} ,\label{Xq9}\\
S_{5(6)}(\hb)=\frac{5040}{Y^6}\Bigl[4-\cos Y+\frac{9\sin Y}{Y}
-\frac{24(1-\cos Y)}{Y^2}\Bigr] ,\label{Xq10}\\
W_{5(6)}^{(1)}(z)=\cases{(1-z)^6(1+3z+3z^2)&at $0\leq z\leq1$,\cr 
0&at $z>1$,}\label{Xq11}\\
\mOmega_{5(6)}^{(1)}(0)=\frac{3}{R_0} .\label{Xq12}
\end{eqnarray}

Interested in the cases of $s\leq6$, we will not propose the complete
set of values characteristic of the next example. However, it is worth 
pointing out that the structure of the corresponding spreading function 
becomes of a more complicated form again and is as follows: 
\begin{equation}\label{Xq13}
g_{7(7)}(r)=\frac{45(1-x)^5(1+5x+8x^2)}{4\pi R_0^3} .
\end{equation}
The corresponding concomitant quantities take the form
\begin{eqnarray}
S_{7(7)}(\hb)&=&\frac{75600}{Y^7}\Bigl[-\sin Y+\frac{24-15\cos Y}{Y}
+\frac{87\sin Y}{Y^2}\nonumber\\
&&{}-\frac{192(1-\cos Y)}{Y^3}\Bigr] ,\label{Xq14}\\
\mOmega_{7(7)}^{(1)}(0)&=&\frac{25}{8R_0}
\end{eqnarray}
and are sufficient to describe point-charge lattices at least while 
different spreading functions of form (\ref{Xq13}) do not overlap.

\end{document}